\documentclass[12pt]{article}
\usepackage{epsf,graphics}

\def\gh#1{}
\def\gh#1{{\large\bf $\spadesuit$ #1}}

\newcommand{\beq}{\begin{equation}}
\newcommand{\eeq}{\end{equation}}
\newcommand{\beqa}{\begin{eqnarray}}
\newcommand{\eeqa}{\end{eqnarray}}
\newcommand{\lsim}{\stackrel{<}{_\sim}}

\def\A{{\cal A}}

\def\Ab{\overline{\cal A}}

\def\be{\begin{equation}}
\def\ee{\end{equation}}
\def\bea{\begin{eqnarray}}
\def\eea{\end{eqnarray}}

\def\lsim{\mathrel{\lower4pt\hbox{$\sim$}}\hskip-12pt\raise1.6pt\hbox{$<$}\;
}

\def\xba{\bar}

\begin{document}

\begin{titlepage}

\begin{flushright}
LMU 09/03\\
AMES-HET 03-04\\
July 2003\\
\end{flushright}

\vspace{0.5cm}
\begin{center}
\Large\bf\boldmath
Implications of Non-Standard CP Violation
in Hadronic $B$-Decays
\unboldmath
\end{center}

\vspace{0.8cm}
\begin{center}
David Atwood$^{1*}$\\[0.1cm]
{\sl  Department of Physics and Astronomy, Iowa State University \\
Ames, IA 50011
} \\[0.4cm]

Gudrun Hiller$^2$\\[0.1cm]
{\sl
Ludwig-Maximilians-Universit\"at M\"unchen, Sektion Physik, 
Theresienstra\ss{}e 37 \\ D-80333 M\"unchen, Germany}

\end{center}

\footnotetext[1]{email: atwood@iastate.edu}
\footnotetext[2]{email: hiller@theorie.physik.uni-muenchen.de}

\vspace{0.3cm}
\begin{abstract}
\vspace{0.1cm}
\noindent

We investigate a class of models for new physics which could produce
a large difference in $\sin 2\beta$ between $B^0\to J/\psi K_S$ and the
``pure penguin'' mode $B^0\to \phi K_S$. In such models, the dominant
effect is through a Z-penguin and therefore a pattern of deviation in
$\sin 2\beta$ as measured in $B^0\to \chi_1 K_S$;  $\eta_c K_S$;
$J/\psi K_S$ and $\psi^\prime K_S$ is predicted. If the preliminary data
concerning the discrepancy between $J/\psi K_S$ and $\phi K_S$ proves
correct in magnitude, discrepancies in these other modes would likely be
observable.  We also consider the effects such new physics could have on
the $B_s$ system and the isospin analysis of $B\to K\pi$.  We
compare this scenario with a scenario where contributions to $\sin 2\beta$
in various modes is produced by gluino loops and down squark mixing.

\end{abstract}

\vspace{0.4 cm}

* Work supported by DOE contract DE-FG02-94ER40817.

\end{titlepage}

\section{Introduction}

The Standard Model (SM) provides a consistent explanation of CP violation
which has been observed in the Kaon system and, more recently
at the BaBar and BELLE B-factories. 
Indeed, the B-factories provide a number of decays where CP
violation should be evident and therefore will allow stringent testing of
the SM mechanism of CP violation.

The most well established B-factory result is the determination of
$\sin 2\beta$ from the decay $B\to \psi K_S$ and related processes
 \cite{sin2betababar}-\cite{Nir:2002gu}:
\begin{eqnarray}
\sin 2\beta_{\psi K} = +0.734\pm 0.054
\label{beta_psiks}
\end{eqnarray}
The SM expectation derived from 
the Kaon sector, the rate of $b\to u$ transitions and 
the rate of $B \bar B$ oscillation is that \cite{CKM-fit}:
\begin{eqnarray}
0.64\leq \sin2\beta_{fit}\leq 0.84~~~(95\%~\mbox{C.L.})
\label{beta_fit}
\end{eqnarray}
and so there is excellent agreement with the above result.

A firm prediction of the SM is that the $b\to s$ penguin would have the
same weak phase to order $\lambda^2$ as the $b\to c \xba c s$ transition
driving $B\to \psi K_S$, see \cite{Grossman:1997gr} for
experimental tests of the SM back ground. 
It therefore follows that in the ``pure penguin''
decay $B\to\phi K_S$ the time dependent CP violation would be the same
as in $B\to\psi K_S$. Early results from the B-factories, however, seem to
contradict this.  If we interpret the time dependence of $B\to\phi K_S$
as a measure of $\beta$, $\beta_{\phi K_S}$, the experimental results are:
\begin{eqnarray}
\begin{array}{rcll}
\sin 2 \beta_{\phi K_S} &=& -0.18 \pm 0.51 \pm 0.07 & 
(BaBar \cite{babarphi} )\cr 
\sin 2 \beta_{\phi K_S} &=& -0.73 \pm 0.64 \pm 0.22 & (BELLE \cite{bellephi} ) 
\cr
\sin 2 \beta_{\phi K_S} &=& -0.38 \pm 0.41 & (Average) 
\end{array}
\label{beta_phiks}
\end{eqnarray}
The result Eq.~(\ref{beta_phiks}) provide a suggestive contrast
to $\beta_{\phi K_s}$ in Eq.~(\ref{beta_psiks}).  If this 2.7-sigma
discrepancy proves, with more data, to be a real difference between $\sin
2\beta_{\psi K_S}$ and $\sin 2\beta_{\phi K_S}$ then it would provide
definitive evidence for new physics (NP).

This discrepancy is only possible if the NP contribution to $B\to \phi
K_S$ is comparable to the SM. This places considerable
constraints on the nature of the NP. In~\cite{gudrun} it was argued that a
promising class of models that can explain the effect involve flavor
changing couplings of the Z-boson to $\bar b s$. Such couplings may arise
either from the $d$-quark mass matrix if these quarks are mixed with extra
vector-like down quark or through flavor changing penguin graphs
containing new particles, such as the
minimal supersymmetric Standard Model (MSSM) with large mixing in the 
up squark sector. Regardless of how the
$s Z b$ coupling arise, this mechanism leads to a number of definite
predictions. 
In this paper, we will discuss some of the implications of non-standard
$sZb$ couplings.
In particular the same amplitude which gives rise to the $\phi K_S$ anomaly 
should also contribute to related hadronic decays $B \to K \pi$;
and indeed to the golden modes such as $\psi K_S$ and $\chi_{c} K_S$. 
The pattern of contributions
to these case may provide additional evidence for this mechanism, and in 
general for any kind of physics beyond the SM.

While NP effects in various $b \to q_1 \bar q_2 q_3 $ decays have been 
estimated e.g.~\cite{Grossman:1996ke} 
we investigate here differences in the CP asymmetry 
induced by interference between 
mixing and decay among different final states with the {\it same} 
flavor content. If NP breaks parity and CP, we obtain in general
different answers from mesons with different 
$J^{CP}$ quantum numbers.

In the case of 
$B \to (c \bar c) K_S$  
where $(c \bar c)=J/\psi, \psi^\prime,\eta_c, \chi_{1}$ 
early experimental analysis 
have already 
been performed as part of the effort to determine $\sin 2\beta$ in the
context of the SM. 
Their weighted average, in fact,  enters the number given in
Eq.~(\ref{beta_psiks}) as
\footnote{Also Belle \cite{sin2betabelle} 
uses $J/\Psi,\Psi^\prime,\eta_c, \chi_{1}$ in their
analysis, however, do not give the individual contributions.}

\begin{eqnarray}
\sin 2 \beta_{J\Psi K_S (K_S \to \pi^+ \pi^-)} 
&=& 0.82 \pm 0.08  \nonumber \\ 
\sin 2 \beta_{\Psi^\prime K_S (K_S \to \pi^+ \pi^-)} 
&=& 0.69 \pm 0.24  \nonumber  \\
\sin 2 \beta_{\chi_1 K_S} &=& 1.01 \pm 0.40  ~~~~~~~~~~~  
(BaBar \cite{sin2betababar})
 \label{eq:ccbardata} \\
\sin 2 \beta_{\eta_c  K_S} &=& 0.59 \pm 0.32  \nonumber
\end{eqnarray}

\noindent Clearly at the present time $\sin 2\beta$ is consistent between
these modes but it will be important to pursue further experimental study
sensitive to the level at which NP enters the $b \to c \bar c s$ decay
amplitude as well as in the corresponding $B_s$-decays. Such studies can
be done at the B-factories and/or at future hadron colliders. It is
important to keep in mind that these searches for NP are based on the
prediction of the SM that $\sin 2\beta$ is consistent for all such modes
and is independent of fits to the unitarity triangle of $\sin 2 \beta$
which are currently dominated by a theory error of $14 \%$ that exceeds
the experimental error in $\sin 2\beta$ by a factor of two, see
Eqs.~(\ref{beta_psiks}) and (\ref{beta_fit}).

The outline of the paper is as follows. In Section~\ref{sec:NPinphi} we
work out constraints on NP from $B \to \phi K_S$ data. In
Section~\ref{sec:general} the effect of non-standard $Z$-couplings in
hadronic 2-body decays are calculated and constraints are discussed.  In
Section~\ref{sfive} we compare the predictions of the scenario with non-SM
Z-penguins with other models of NP and in Section~\ref{ssix} we conclude.  
In Appendix \ref{app:heff} we give the matrix elements of $\bar B \to \phi
K$ and $\bar B \to (c \bar c) K$ decays in terms of the effective low
energy Hamiltonian. In Appendix \ref{app:Zp} the initial conditions and
the leading log running of the ElectroWeak penguins are specified.

\section{The $B \to \phi K_S$ decay amplitude with NP \label{sec:NPinphi}}

Let us now assume that $B\to \phi K_S$ receives a NP amplitude
which accounts for the discrepancy in $\sin 2\beta_{\phi K_s}$.
Thus, if $a$ is the magnitude of the SM contribution and $b$
is the magnitude of the NP contribution, then if we use a phase
convention where the SM contribution is real the amplitude may be written 
as:

\begin{eqnarray}
\A (\phi K_s)&=&a+b e^{i(\rho+\lambda)}
\nonumber\\
\Ab(\phi K_s)&=&\eta(X) \left [a+b e^{i(\rho-\lambda)} \right ]
\label{decomp}
\end{eqnarray}

\noindent
where $\rho (\lambda)$ is the strong (weak) phase difference 
between the two contributions and 
$\A(X)$ is the amplitude for $B^0\to X$ while
$\Ab(X)$ denotes the amplitude for $\bar B^0\to X$.
The factor $\eta(X)$ denotes the CP eigenvalue of the final state $X$, 
e.g.~$\eta(\phi K_S)$=-1.

It follows then that 
the time dependent CP asymmetry may be written as:
\begin{eqnarray}
\label{eq:acpt}
a_{CP}(t)=\frac{\Gamma(\bar B^0(t)\to X)-\Gamma(B^0(t)\to X)}
{\Gamma(\bar B^0(t)\to X)+\Gamma(B^0(t)\to X)}=
-C_X \cos(\Delta m_B t) +S_X \sin(\Delta m_B t)
\end{eqnarray}
where the sine and cosine coefficients are given by:
\begin{eqnarray}
C_X&=&
{|\A(X)|^2-|\Ab(X)|^2\over |\A(X)|^2+|\Ab(X)|^2}\nonumber\\
S_X&=&
{2 Im(\A^*(X)\Ab(X) e^{-2i\beta})   \over |\A(X)|^2+|\Ab(X)|^2}
\equiv -\eta(X) \sin 2 \beta_X
\label{eq:SX}
\end{eqnarray}
Here, $2 \beta$ is the
phase from the $B \bar B$ mixing
defined in the phase convention of Eq.(\ref{decomp})
\footnote{We comment on
NP in $B \bar B$ mixing in Section \ref{ssix}.}.

%
%
%
%
%
%
%

From the observed time dependent CP violation of $B\to \phi K_S$ we may
obtain some information about these amplitudes on a model independent
basis.  In the following we will discuss the analysis of the data in the
scenarios that the strong phase difference $\rho=0$.

In the case where there is no strong phase, then $\A^* \eta(X)=\Ab$ and 
therefore
$C_X=0$. We can therefore extract the amplitude from the CP averaged 
branching ratio  $\Gamma_0={1\over 2}(|\A|^2+|\Ab|^2)$ and
$S_X$ using:
\begin{eqnarray}
\A(X)=\sigma_1  
\left [
\sqrt{\Gamma_0(X)}
\sqrt
{
-i \eta(X) S_X
+
\sigma_2
\sqrt{1-S_X^2}
}
\right ]
e^{-i\beta}
\label{no_st_ph}
\end{eqnarray}
where $\sigma_{1,2}=\pm 1$ giving a 4-fold ambiguity. 
Then, given $\beta$ 
we can determine from Eq.~(\ref{decomp}) for each value of $a$ a 
value of $b/a$ and $\lambda$ by $b/a \; e^{i\lambda}=\A/a-1$.

%
%
%
%
%
%
%
%
%
%
%
%
%
%
%

\noindent
Data are consistent with no direct CP violation in $B \to \phi K_S$ decay:
\begin{eqnarray}
\begin{array}{rcll}
C_{\phi K_S} &=& -0.80 \pm 0.38 \pm 0.12 & 
(BaBar \cite{babarphi} )\cr 
C_{\phi K_S} &=& +0.56 \pm 0.41 \pm 0.16 & (BELLE \cite{bellephi} ) 
\cr
C_{\phi K_S} &=& -0.19 \pm 0.30 & (Average) 
\end{array}
\end{eqnarray}

\begin{figure}
\begin{center}
\epsfxsize 4.6 in
\mbox{\epsfbox{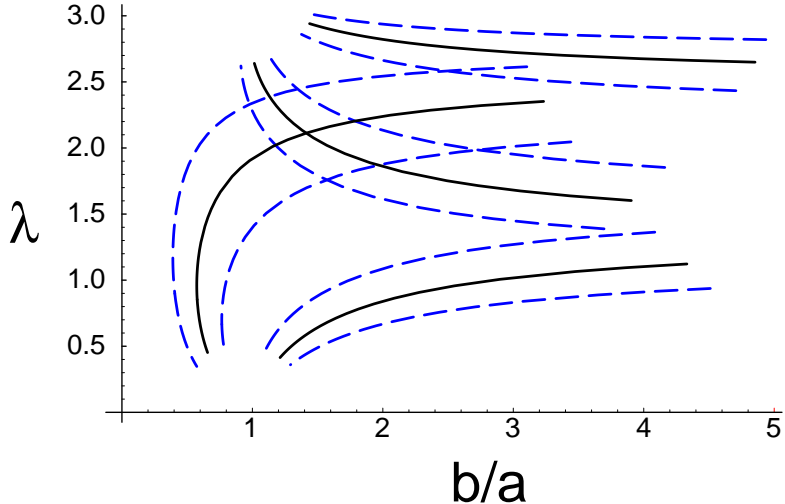}}
\caption{
4-fold solutions following from 
Eq.~(\ref{no_st_ph}) in the 
$b/a$ - $\lambda$ plane for  $S_{\phi K_S}=-0.79,-0.38,+0.03$, i.e.~the
central experimental value and the 1 $\sigma$ range 
for vanishing strong phase. 
The curves corresponding to the central value (-0.38) are solid while the
curves
corresponding to the 1-sigma range (-0.79,+0.03) are dashed.
For details see text.
}
\label{scat_1}
\end{center}
\end{figure}

In Fig.~\ref{scat_1} we show 
the weak phase $\lambda$ as a function of the NP
to SM ratio $b/a$ for the case without a strong phase. 
The curves are the solutions to Eq.~(\ref{no_st_ph})
for  the central value and 
the $\pm$ 1 sigma range of $S_{\phi K_S}(=\sin 2 \beta_{\phi K_S})$
from  the data average  
in Eq.~(\ref{beta_phiks}) and $\beta =23.9^\circ$, the central value of the
fit  Eq.~(\ref{beta_fit}).
The measured branching ratio 
${\cal{B}}(B^0 \to \phi K^0)=8.4 \pm 1.6 \cdot 10^{-6}$ \cite{PDG2002}
agrees within errors with the SM value, e.g.~\cite{Chiang:2003jn}.
To avoid a ``fine tuning'' configuration i.e.~a near cancellation between NP
and SM we take 
$\sqrt{\Gamma_0 (\phi K_S)}/4 \leq a\leq 2 \sqrt{\Gamma_0 (\phi K_S)}$.
We recall that for fixed input $S,\Gamma_0$ each value of 
$b/a$ gives a 4-fold solution for $\lambda$ within $0$ and $\pi$.
Some solutions in Fig.~\ref{scat_1} end for larger values of $b/a$ because 
of the lower bound on $a$.
The cut-out regions around 
$b/a \sim 1$ increase in size if the upper bound
on $a$ would decrease and vice versa.
We see that ratios $b/a$ are typically order one with order one weak phase
$\lambda$, similar to the findings of Ref.~\cite{Chiang:2003jn}.

\section{$Z$-penguin effects in 2-body $b$-decays  \label{sec:general}}

The Lagrangian of the effective FCNC $sZb$ couplings maybe written as
\begin{eqnarray}
{\cal{L}}_{Z}=\frac{g^2}{4 \pi^2} \frac{g}{2 \cos \theta_W}
\left( \bar b_L \gamma_\mu s_L Z_{sb}+\bar b_R \gamma_\mu s_R Z^\prime_{sb}
\right) Z^\mu+ h.c.
\end{eqnarray}
where $Z_{sb}$ ($Z^\prime_{sb}$) denote the left (right) handed coupling
strength.
The $sZb$-couplings are experimentally 
constrained as 
\begin{eqnarray}
\label{eq:bound}
\sqrt{|Z_{sb}+Z_{sb \; SM}|^2+|Z_{sb}^\prime|^2} \leq 0.08
\end{eqnarray}
which updates \cite{Ali:1999mm,GGG}, where details can be found.
The bound in Eq.~(\ref{eq:bound}) is based on inclusive $B \to X_s e^+ e^-$
decays at NNLO \cite{aghl} 
and corresponds to an enhancement of 
$2-3$ over the SM value 
\footnote{The experimental bounds on the 
branching ratios of $B \to (X_s,K,K^*) \ell^+ \ell^-$ decays have gone down, 
but the theoretical value has decreased from NLO to NNLO, too.}
\begin{equation}
Z_{sb \; SM}= -V_{tb}^* V_{ts} \sin^2 \theta_W C_{10}^* \simeq -0.04 \; , 
~~~~~~~~~~ Z_{sb \; SM}^\prime \simeq 0
\end{equation}
The inclusion of right handed couplings into the distributions of semileptonic 
$b \to s \ell^+ \ell^-$ decay is straightforward  since different helicities
do not interfere in the limit of a vanishing strange quark mass.

Tree level $Z$-boson exchange and subsequent $q  \bar q$
pair production induces NP contributions to hadronic $B$-decays
via $b \to  s  \bar q q$. 
These are proportional to the $q \bar q Z$ coupling, which we assume to be 
SM like, i.e., the coupling $Z_V$ to vector and axial vector $Z_A$ currents 
are given, respectively, as
$Z_V=I_{3q}-2 Q_q \sin^2 \theta_W$ and $Z_A=-I_{3q}$. 
Note that we use 
$Z_V  \bar q \gamma_\mu q + 
Z_A  \bar q \gamma_\mu \gamma_5 q $
such that the sign of $Z_A$ is opposite to the PDG definition
\cite{PDG2002}.
The values and ratio of  different $q \bar q Z$ coupling
are compiled in Table \ref{tab:AoverV} for $\sin^2 \theta_W=0.23$.
\begin{table}[htb]
 \begin{center}
\begin{tabular}{|c|c|c|c|}
\hline
$\mbox{}$&$c \bar c$&$s \bar s$& $u \bar u-d \bar d$\\
\hline
$Z_V $   &   +0.19 & -0.35 & +0.54    \\ 
$Z_A $   &   -0.5 & +0.5 & -1      \\ 
$Z_A/Z_V$   &   -2.6 & -1.4 & -1.9     \\ 
\hline
\end{tabular}
\caption{
Values of vector, axial vector $Z$-couplings and their ratio for 
different $q \bar q$ pairs.
}\label{tab:AoverV}
 \end{center}
\end{table}

As can be seen from the ratio $Z_A/Z_V$ 
in Table \ref{tab:AoverV} the $c \bar c$ system has the biggest
spread among final states with different quantum 
numbers (vector versus axial vector coupling) 
induced by the $Z$-exchange.
In the next Sections \ref{subsec:charm} and \ref{sec:Bs}
we estimate the implications
of large, in general CP violating $sZb$-couplings for 
$B \to (c \bar c) K$ and $B_s \to (c \bar c ) \phi$ decays.
Further, the couplings of  $Z$-penguins to $I=1$ mesons such as $\pi^0$
are very large.
In Section \ref{sec:Kpi} we investigate whether current data on 
$B \to K \pi^0$ decays yield additional constraints on the $sZb$-couplings.
Note that in order to explain a deviation in CP asymmetry
from the SM as large as hinted by  current data 
the couplings $Z_{sb}^{(\prime)}$
need to be near their upper bound and with a large phase.

\subsection{$B$-decays into charmonium \label{subsec:charm}}

We discuss the implications of non-standard $Z$-penguins in the 
decays $B \to M K$ into
charmonium $M=\eta_c,\Psi$, $\Psi^{\prime \ldots}$, 
$\chi_0,\chi_1,\chi_2$. The NP effect in the decays 
is split according to the
CP properties of the final $(c \bar c)$ states. Whereas the vector mesons 
couple to a vector current $\sim Z_V$,
the pseudoscalar and axial vector mesons couple to
axial vector current $\sim Z_A$.
Hence, the $Z$-penguin effect in 
$\eta_c$ and $\chi_1$
is bigger than in the vector mesons $\Psi, \Psi^{\prime \ldots}$ by a 
factor of 2.6, as can be read off Table \ref{tab:AoverV}.
In the presence of a CP violating phase in the $Z$-contribution, this
leads to a difference in $\sin 2 \beta$ measured among the {\it golden} modes.
The generic size of this effect is 
$\arg(P/T) P/T$
where $P$ denotes the 
penguin contribution including the NP, and $T$ is the SM 
tree contribution in the $b \to c \bar c s$ 
amplitude. In the SM, $\arg(P/T) \sim \lambda^2$, whereas it can be 
order one in the presence of NP. 
Since ${\cal{B}}(B \to \phi K)/{\cal{B}}(B \to (c \bar c) K) \simeq 10^{-2}$
an order one NP effect in $b \to s \bar s s $ gives up to
ten percent contribution to the $b \to c \bar c s$ amplitude.
Therefore, we expect here differences in $\sin 2 \beta$ of this order.

\begin{figure}
\begin{center}
\epsfxsize=0.7\textwidth
\rotatebox{270}{\epsffile{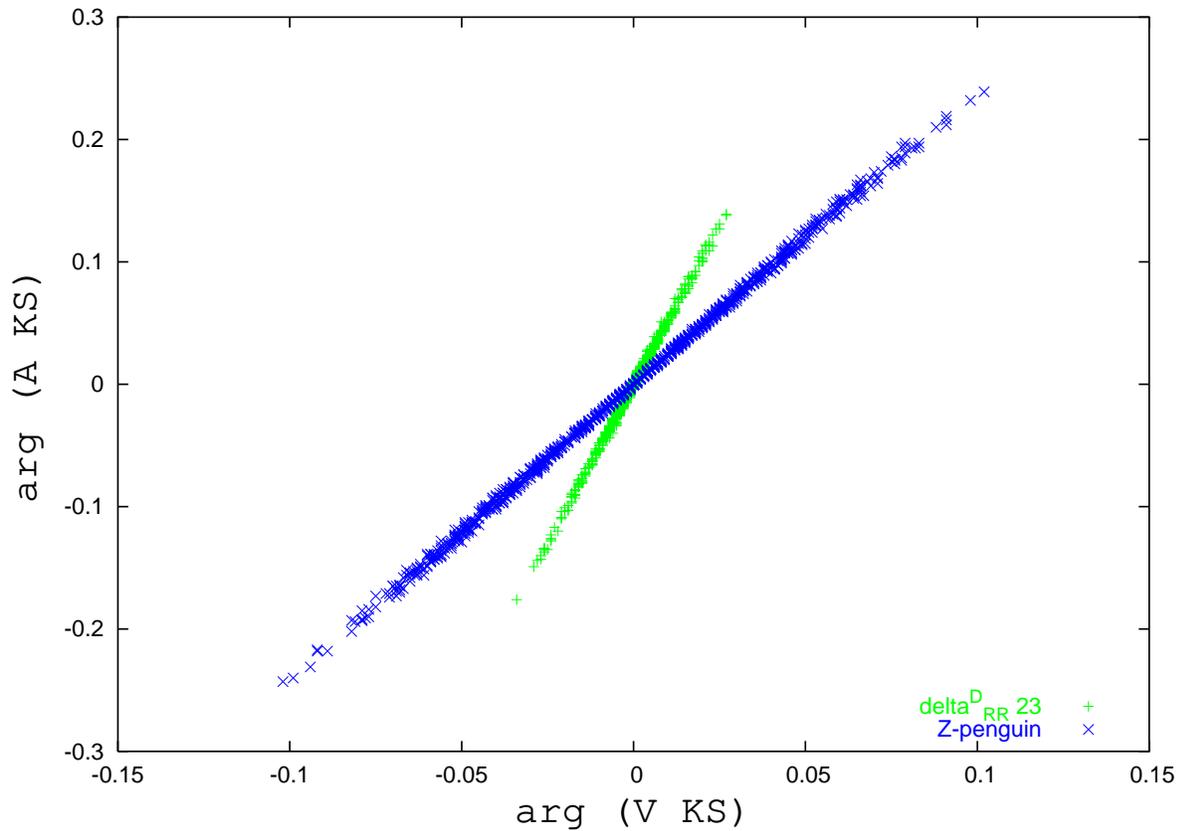}}
\end{center}
\caption{The NP correction for axial coupling mesons
$\arg (A K_S)$ as a function of the one for vector coupling mesons
$\arg (V K_S)$ induced by non-standard $Z$-penguins ($\times$, blue) and in 
the MSSM with down squark mixing $\delta^D_{RR \, 23}$ (+, green) discussed in 
Section \ref{sec:gluino}.}
\label{fig:arg}
\end{figure}

To quantify this general prediction, we 
neglect for simplicity direct CP violation and obtain
for the size of the NP effect
\begin{equation}
\sin 2\beta_{M K_S}- \sin 2\beta = 
\sin (2\beta - \arg (M K_S) )
- \sin 2\beta \approx -\arg (M K_S) \cos 2 \beta
\end{equation}
where we expanded in small phases in $\bar {\cal{A}}/{\cal{A}}$
and abbreviated
$\arg \bar {\cal{A}}( \overline{M K_S})/{\cal{A}}(M K_S) \equiv \arg (M K_S)$.
Note that $\eta(\chi_{0,2} K_S)=+1$ and -1 for
$J/\Psi, \Psi^{\prime \ldots},\eta_c,\chi_1$.
Numerically, we obtain in the non-standard $Z$-scenario
\begin{eqnarray}
 | \arg (M K_S) | \leq
\begin{array}{ll}
 0.11   & ~~~~~ \mbox{for} \; M=\Psi, \Psi^{\prime \ldots} \\
 0.25   & ~~~~~ \mbox{for} \; M=\eta_c,\chi_1 \\
\end{array}
\label{eq:argMK}
\end{eqnarray}
Here we varied the magnitudes and phases 
of $Z_{sb}$ and $Z_{sb}^\prime$ independently while respecting the constraint
given in Eq.~(\ref{eq:bound}).
These numbers as well as numerical estimates of other quantities to be
discussed later in the text
(also for the case with $Z_{sb}^\prime=0$) are summarized in Table 
\ref{tab:Zreach}.
Explicit formulae of the matrix elements and details 
can be seen in Appendix \ref{app:heff} and \ref{app:Zp}.
Since
\begin{eqnarray}
\frac{\arg (A K_S)}{\arg (V K_S)}
\simeq \frac{Im(C_3^{ Z}-C_7^{ Z}+C_9^{ Z}+C_5^{\prime Z}+C_7^{\prime Z}
-C_9^{\prime Z})}
{Im(C_3^{ Z}+C_7^{ Z}+C_9^{ Z}+C_5^{\prime Z}+C_7^{\prime Z}+C_9^{\prime
Z})}
\approx -\frac{Z_A}{Z_V}= +2.6
\end{eqnarray}

\noindent
the NP correction from $Z$-penguins has the same sign
for vector mesons $V=J/\Psi, \Psi^\prime$ as for the axial ones
$A=\chi_1,\eta_c$.
This correlation holds over the whole $(Z_{sb},Z_{sb}^\prime)$ parameter 
space after performing leading log QCD corrections.
This is illustrated in Fig.~\ref{fig:arg}, where the NP phases 
$\arg (A K_S)$ versus $\arg (V K_S)$ are shown for $Z$-penguins 
($\times$, blue).
Also shown is the correlation in another NP scenario (+, green), the MSSM
with additional flavor and CP violation in singlet down squark mixing, 
to be discussed in Section \ref{sec:gluino}.

\noindent
We estimate the $Z$-exchange effect in the difference between 
decays into axial and vector 
coupling charmonia as
\begin{eqnarray}
|\sin 2\beta_{A K_S}-\sin 2\beta_{V K_S} | 
& \leq & 0.18 \cos 2 \beta  
\label{eq:Zprediction}
\end{eqnarray}
Available data given in Eq.~(\ref{eq:ccbardata})
are not significant yet i.e.~the difference of the error weighted 
average of axial minus vector coupling final states reads as  
$\sin 2\beta_{A K_S}-\sin 2\beta_{V K_S} =-0.05 \pm 0.26$.

Decays into  $\chi_{0,2}$ final states are factorization forbidden modes
since by $C$ conjugation and Lorentz invariance  
\begin{eqnarray}
\label{eq:zero}
\langle \chi_0(0^{++}) | \bar c \gamma_{\mu} c         | 0 \rangle =0, ~~~~
\langle \chi_2(2^{++}) | \bar c \gamma_{\mu} c         | 0 \rangle =0
\end{eqnarray}
and their amplitude
${\cal{A}}(\bar B^0 \to \chi_{0,2} \bar K^0)$ requires gluon exchange.
At order $\alpha_s$ the color suppressed $Z$-penguins compete with
a color enhanced SM contribution from tree level $W$ exchange, thus
they are doubly $1/N_C$ 
suppressed with respect to the factorization allowed modes.
Therefore, the effect of NP from $Z$-penguins should be least pronounced
in $B \to \chi_{0,2} K_S$ and time dependent CP asymmetries should
return in these modes the least polluted value 
of $\sin 2 \beta$ in the $c \bar c$ system.
However, since naive factorization is badly broken in these modes, it is 
difficult to be quantitative here. Still, a departure in  $\sin 2 \beta$ 
in $\chi_{0,2}$ from the fitted value given in Eq.~(\ref{beta_fit}) 
indicates the presence of NP, as well as any discrepancy in the extracted 
value of  $\sin 2 \beta$ among different charmonia.

\subsection{$Z$-penguin effects in the $B_s$-system \label{sec:Bs}}

We investigate here implications on 2-body decays in the $B_s$-system such as
$B_s \to (c \bar c) \phi$ and $B_s \to \phi \phi$.
The coefficient of the $\sin (\Delta m_{B_s} t)$ term in the time dependent 
CP asymmetry reads as (see Eqs.~(\ref{eq:acpt}) and (\ref{eq:SX}) with
changes from  $B_d$ to $B_s$-mesons) 
\begin{eqnarray}
S_{M \phi} = -\eta(M \phi) 
\sin \left (\arg M_{12} - 
\arg (M K_S) \right)
\end{eqnarray}
where we neglected for simplicity the width difference between $B_s$ and 
$\bar B_s$ mesons and as before direct CP violation, i.e. we used 
$|\bar {\cal{A}}/{\cal{A}}|=1$.
In the SM the phase of the mixing amplitude 
$\arg M_{12}^{SM}=-2 \beta_s=-2 \lambda^2 \eta$
is tiny. Hence, we expect $S_{c \bar c \phi}$
and  $S_{\phi \phi}$ of ${\cal{O}}(\lambda^2)$ in the SM.
The
$sZb$-penguins do have a twofold effect in the $B_s$-system, on the decay 
amplitude similarly to the $B_d$-system and on the $B_s \bar B_s$ mixing.
Hence, both $\Delta B=1$ and  $\Delta B=2$ terms 
in $S_{M \phi}$ receive contributions from the NP.
Concerning the latter
and employing the notations of \cite{BBL}, we find
\begin{eqnarray}
M_{12}^Z =\frac{\alpha G_F^2 m_W^2}
{3 \pi^3 \sin ^2 \theta_W} B_{B_s} f_{B_s}^2 m_{B_s} \eta_B
\left(Z_{sb}^{*2}+ Z_{sb}^{\prime * 2} +2 Z_{sb}^* Z_{sb}^{\prime *} X \right)
\end{eqnarray}
Here, $X=4 \bar P_1^{LR}=-2.84$ \cite{Buras:2002vd}
includes differences in the bag factors (taken from lattice \cite{lattice}) 
and perturbative QCD corrections of the
matrix element between operators with different Dirac structure, 
$ \gamma_\mu L(R) \times \gamma^\mu L(R)$ and 
$\gamma_\mu L(R) \times \gamma^\mu R(L)$.
We find 
\begin{eqnarray}
r_Z  \equiv 
\frac{M_{12}^Z}{M_{12}^{SM}} = 
\frac{ 4 \alpha}{\pi \sin^2\theta_W S_0(x_t) }
\left(\frac{Z_{sb}^{*2}+ Z_{sb}^{\prime * 2} +2 Z_{sb}^* Z_{sb}^{\prime *} X}
{(V_{tb} V_{ts}^*)^2}\right) 
\label{eq:BBmix}
\end{eqnarray}
a correction of up to $|r_Z| \leq 0.5$  w.r.t.~the SM.

The analysis of NP on the decay amplitude 
is analogous to the discussion in
Section \ref{subsec:charm}.
However because the final state in these modes consists of two mesons 
of spin$\neq 0$, to  
isolate the CP eigenstate components of the final state 
one has to perform angular analysis
in $B_s \to M \phi$ decays with
$M=J/\Psi,\Psi^\prime,\chi_{1,2},\phi$. Otherwise, 
there is a dilution in $S_{M \phi}$ from admixture of CP odd and even 
contributions which diminishes a potential NP effect  and
introduces an additional uncertainty. 
None the less, 
this is an excellent null test of the SM since
any large CP asymmetry, even in the data summed over
polarization, would indicate the presence of NP.
If the particle recoiling against the $\phi$ is a scalar/pseudoscalar,
then of course there is a single amplitude and 
angular analysis is not required. This would be the case for 
$\chi_0$ and $\eta_c$.

With $\arg M_{12}=\arg(1+r_Z)$ we find 
\begin{eqnarray} 
|S_{M \phi}| & \leq &
\begin{array}{ll}
 0.42 & ~~~~~ \mbox{for} \; M=\Psi, \Psi^{\prime \ldots} \\
 0.47 & ~~~~~ \mbox{for} \; M=\eta_c,\chi_1 \\
 0.66 & ~~~~~ \mbox{for} \; M=\phi \\
\end{array}
\end{eqnarray}
The spread in $S_{M \phi}$ between different charmonia is up to 0.15.
The corresponding numbers for the case with the flipped  helicity 
$Z$-coupling $Z_{sb}^\prime$ switched off for all quantities discussed 
in this section can be seen in Table \ref{tab:Zreach}.

\subsection{Isospin analysis in $B \to K \pi$ decays \label{sec:Kpi}}

\begin{figure}
\begin{center}
\epsfxsize=0.33\textwidth
\rotatebox{270}{\epsffile{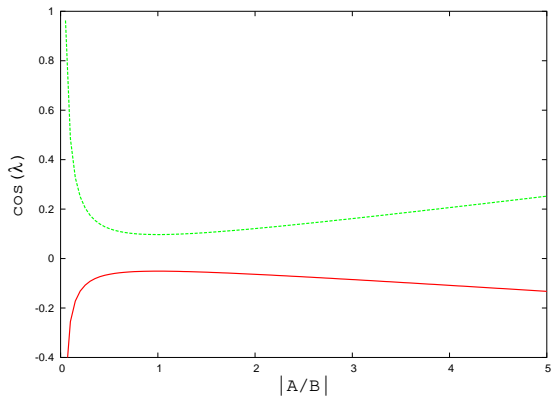}}
\end{center}
\caption{Constraints on isospin violation from $B \to K \pi^0$ data for
vanishing strong phases. The bounds weaken for non-zero strong phases.
The range between the curves is allowed at 1 $\sigma$.}
\label{fig:pi0}
\end{figure}

We start with an analysis in  $B \to K \pi^0$ decays. 
Isospin analysis relates the decay amplitudes of neutral and charged $B$ mesons
as \cite{Gronau:1991dq}
\begin{eqnarray}
{\cal{A}}(B^0 \to K^0 \pi^0) = B-A, ~~~~
-{\cal{A}}(B^+ \to K^+ \pi^0) = B+A
\end{eqnarray}
where 
$B\equiv B_{1/2}$ denotes the $\triangle I=0$ piece and
$A\equiv A_{1/2}-2 A_{3/2}$ is a $\triangle I=1$ combination 
of $I_f=1/2$ and $I_f=3/2$.
$Z$-penguins do violate isospin and hence will give a significant contribution
to $A$. Is the amount required to explain the anomaly in
$B \to \phi K_S$ i.e.~of 
size comparable to the $\triangle I=0$ SM QCD penguins
allowed by current data ? 
Experimental findings \cite{recentdata}
give 
\begin{eqnarray}
r \equiv  
\left( \frac{({\cal{B}}(B^+ \to K^+ \pi^0) +{\cal{B}}(B^- \to K^- \pi^0))
\tau(B^0)}
{({\cal{B}}(B^0 \to K^0 \pi^0)+{\cal{B}}(\bar B^0 \to \bar K^0 \pi^0)) 
\tau(B^+)} \right)^{1/2}=1.03 \pm 0.08
\end{eqnarray}
This can be written in terms of a relative strong (weak) phase 
$\rho(\lambda)$  and
ratio $|A/B|$ with relation
\begin{eqnarray}
\cos \lambda \cos \rho =- \frac{1-r^2}{1+r^2} \frac{1+|A/B|^2}{2 |A/B|}
\end{eqnarray}
The 1 $\sigma$ 
allowed $\cos \lambda,|A/B|$ parameter space 
for vanishing strong phases is shown in Figure \ref{fig:pi0}.
For non-zero $\rho$ the allowed range increases and the constraint 
disappears for $\cos \rho =0$.
Therefore, large isospin breaking contributions $A/B \sim O(1)$
are currently not constrained by $B \to K \pi^0$ data 
if the phase $\lambda\sim O(1)$ is large, 
if, for example, the NP contribution comes with a non CKM large weak phase.

Another constraint on non-standard $Z$-penguins can 
come from $B \to \pi \pi$,
$B \to K \pi$ data and $SU(3)$ symmetry.
These Neubert-Rosner type bounds constrain the ratio

\begin{eqnarray}
R_\ast^{-1} =\frac{2 \left[{\cal{B}}(B^+ \to \pi^0 K^+)+ 
{\cal{B}}(B^- \to \pi^0 K^-)\right] }
{{\cal{B}}(B^+ \to \pi^+ K^0)+ {\cal{B}}(B^- \to \pi^- \bar K^0)}
\end{eqnarray}

\noindent
in the presence of isospin and CP breaking NP by
\begin{eqnarray}
\label{eq:boundR}
R_\ast^{-1} \geq
\left[ 1- \bar \epsilon_{3/2} \sqrt{(|a|+|\cos \gamma|)^2+
(|b|+|\sin \gamma|)^2 }\right]^2 
\end{eqnarray}
where $a,b$ are CP  even (odd) isospin violating contributions, see
\cite{Grossman:1999av} for details.
There is a corresponding upper bound on $R_\ast^{-1}$ obtained 
by interchanging the sign in front of the $\bar \epsilon_{3/2}$ term.
As discussed in Section \ref{sec:general}, 
semileptonic rare $b \to s \ell^+ \ell^-$ decays bound enhanced
$Z$-penguins to be at most 2 to 3 of their value in the SM, 
i.e.~$|a + i b| \lsim (2-3) |a_{SM}|$.
Hence, if
the SM is allowed, then the upper bound on $R_\ast^{-1}$ is even 
less constraining for $Z$-penguins. However, the lower bound 
excludes large values of $|a|,|b|$, as can be seen from Eq.~(\ref{eq:boundR}).
For our analysis we find  $R_\ast^{-1}=1.31 \pm 0.15$, 
$\bar \epsilon_{3/2}=0.20 \pm 0.02$ using 
recent data \cite{recentdata} and
use $a_{SM}=0.64$, $b_{SM}=0$ \cite{Grossman:1999av} for the central SM value.
\begin{figure}
\begin{center}
\epsfxsize=0.45\textwidth
\rotatebox{0}{\epsffile{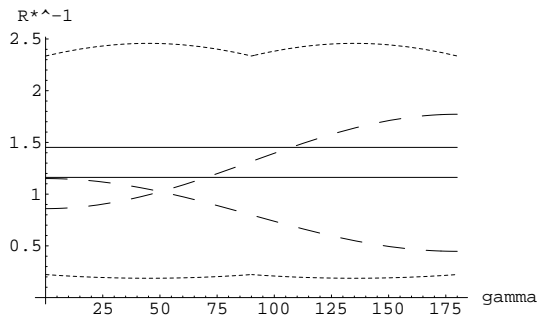}}
\end{center}
\caption{The ratio $R_\ast^{-1}$ as a function of $\gamma$.
Shown are the data at 1 $\sigma$ (solid), 
the bounds in a NP
scenario with enhanced and order one CP violating $Z$-penguins (dotted) and
the SM bounds 
for central values (dashed).}
\label{fig:Rstar}
\end{figure}

We plot the ratio $R_\ast^{-1}$as a function of $\gamma$
in Fig. \ref{fig:Rstar}. Displayed are the 1 $\sigma$ band from data (solid),
the NP bounds with $|a|=|b|=2 |a_{SM}|$ (dotted) and
for comparison
the SM bounds (see \cite{Grossman:1999av}) using central values (dashed).
We see that for such moderate
enhancement of the isospin violating contributions 
the NP bounds are even weakened compared to the SM, 
hence irrelevant.
Note that in order to accommodate 
$\sin 2 \beta_{\phi K_S}=-0.4$, one needs order one CP violation, 
i.e.~$|a| \sim |b|$.
We checked that our conclusions are stable under variation of the
sizes of $a$,$b$ independently as long as $|a|,|b| \leq {\cal{O}}(10)$.

\section{Predictions of $Z$-penguins vs other NP models \label{sfive} }

In Sections \ref{subsec:charm} and \ref{sec:Bs} we have discussed the impact of
complex, non-standard $Z$-penguins on time dependent asymmetry measurements
$B \to (c \bar c ) K$ and $B_s \to (c \bar c ) \phi$ decays.
In this Section we discuss further predictions of this NP scenario and 
compare them with those of other models.

Non-standard $Z$-penguin effects 
in $B \to \phi K_S$ have been first discussed in \cite{gudrun}.
Unlike $B \to (c \bar c) K$ decays where the chromomagnetic dipole operator 
is color octet and suppressed, NP in $b \to s \bar s s$ decays 
can enter in both the 4-Fermi and the gluon dipole operators 
$O_{8 g}^{(\prime)}$, see Appendix 
\ref{app:heff} and \ref{app:Zp}.
We stress that the NP in a scenario with non standard $Z$-penguins 
is confined to the 4-Fermi 
operators, i.e.~dominantly in the electroweak penguin $O_9$.
We checked that the NP effects induced in the dipole operators via
leading log RGE mixing are below few percent.
This is good since there are substantial theoretical uncertainties related
to the matrix element of $O_{8 g}^{(\prime)}$.
Therefore, 
if it turns out that there is NP in $B \to \phi K_S$ decay but not
in $b \to c \bar c s$ - something that can be checked by comparing
$\sin 2 \beta$ from decays into different charmonia - the scenario with 
non-standard $sZb$-couplings - and any other who does not have CP violation
beyond the SM in the chromomagnetic dipole operator - can be excluded.

\begin{figure}
\begin{center}
\epsfxsize=0.5\textwidth
\rotatebox{270}{\epsffile{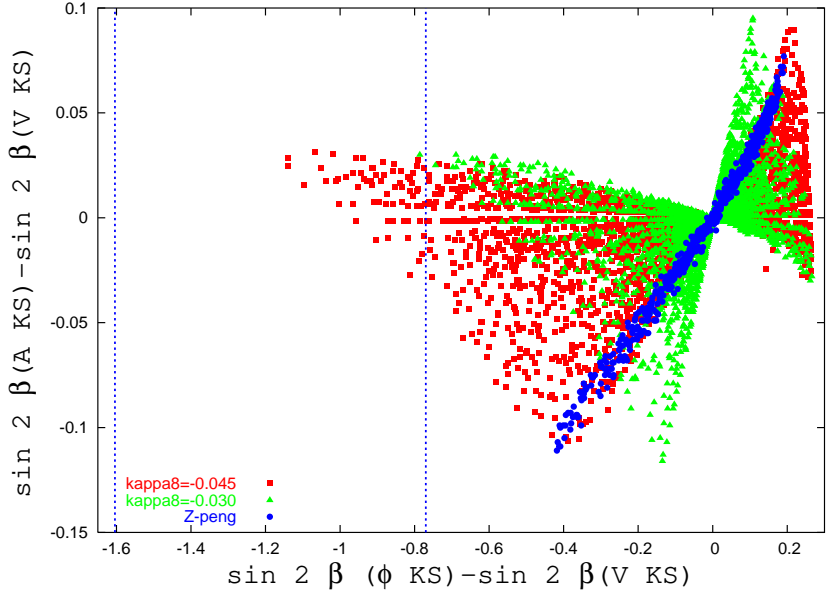}}
\end{center}
\caption{Difference in mixing $\left[ \sin 2\beta_{A K_S}-
\sin 2\beta_{V K_S} \right]$ as a function of
$\left[ \sin 2\beta_{\phi K_S}-\sin 2\beta_{V K_S} \right]$
in the non-SM $Z$-scenario (blue) and in the MSSM with additional 
flavor violation induced by $\delta^D_{RR \; 23}$.
The latter is shown for $\kappa_8=-0.045$ (red) and $\kappa_8=-0.030$ (green).
Also displayed is the 1 $\sigma$ range from data given in 
Eq.(\ref{eq:diffdata}).}
\label{fig:phiccbar}
\end{figure}

The correlation of CP asymmetries in $B \to \phi K_S$ 
with decays into charmonium
with axial $A=\chi_1, \eta_c$ and and vector coupling $V=J/\Psi, \Psi^\prime$
is shown in Fig.~\ref{fig:phiccbar}. To be specific, we plot
$\sin 2\beta_{A K_S}-\sin 2\beta_{V K_S}$ as a function of
$\sin 2\beta_{\phi K_S}-\sin 2\beta_{V K_S}$
in the $Z$-penguins scenario (blue).
Also shown is the outcome in the MSSM with down squark mixing
$\delta^D_{RR \, 23}$ discussed in Section \ref{sec:gluino}
for two values of the chromomagnetic 
$\alpha_s$ matrix element, $\kappa_8=-0.045$ (red) and 
$\kappa_8=-0.030$ (green).
Data given in Eqs.~(\ref{beta_phiks}) and (\ref{eq:ccbardata}) yield
\begin{eqnarray}
\sin 2\beta_{\phi K_S}-\sin 2\beta_{(J/\Psi, \Psi^\prime) K_S} =-1.19 \pm 0.42
~~~~~~~~~(Average)
\label{eq:diffdata}
\end{eqnarray}
which is also displayed.
The correlations shown in Fig.~\ref{fig:phiccbar} from the two 
beyond the SM models 
are very distinct and allow to distinguish between them with improved data.

\subsection{Impact of supersymmetric $\tilde g-\tilde d$ loops on
$b \to c \bar c s$ decays \label{sec:gluino} }

Let us now consider 
the implications of the MSSM with additional
flavor and CP violation beyond the Yukawa couplings on $b \to c \bar c s$ 
transitions. In particular, let us suppose that  there is
only mixing between the singlet scalar partners of the 
down quarks of 2nd and 3rd generation, to be abbreviated here as 
$\delta^D_{RR \; 23}$. This term induces FCNC though gluino-down squark loops.
In this scenario, $\delta^D_{RR \; 23}$ 
the only source of 
beyond the SM CP violation and so this particular framework is rather
predictive.
For previous studies of gluino mediated effects in $B \to \phi K_S$ see
\cite{Harnik:2002vs},\cite{other}.

The model gives  contributions to the flipped 
4-Fermi operators $O_{3 \ldots 6}^\prime$ and the dipole operators 
$O_{7 \gamma}^\prime$, $O_{8 g}^\prime$. 
We use the initial conditions
at the weak scale given in \cite{Gabbiani:1996hi} in
the mass insertion approximation. 
We evolve the coefficients
$C_i$ from the SM and $C_i^\prime$ from the NP separately to the
$m_b$ scale with the leading log RGE, see e.g~\cite{BBL}.
We scan over the squark mass
$150 \mbox{GeV} \leq m_{\tilde q} < 1 \mbox{TeV}$, the gluino mass
$0.2  \leq  ( m_{\tilde g}/m_{\tilde q})^2 < 1.3$
and $|\delta^D_{RR \; 23}| \leq 1$ 
\footnote{ At large $\tan \beta$ the magnitude of
$\delta^D_{RR \; 23}$ is constrained by $B_s \to \mu^+ \mu^-$ data 
\cite{Isidori:2002qe}.}
with arbitrary phase and include the 
constraints from data on ${\cal{B}}(b \to s \gamma)$ following \cite{aghl},
\cite{Kagan:1998ym}.
Numerically, we find
\begin{eqnarray}
| \arg (M K_S) | \leq
\begin{array}{ll}
 0.04 & ~~~~~ \mbox{for} \; M=\Psi, \Psi^{\prime \ldots} \\
 0.20 & ~~~~~ \mbox{for} \; M=\eta_c,\chi_1 \\
\end{array}
\label{eq:argMKgluino}
\end{eqnarray}
Similar to $Z$-penguins, the NP effects in decays to 
axial mesons $A=\chi_1, \eta_c$ 
are bigger than in the decays to vector mesons 
$V=J/\Psi, \Psi^\prime$. We find that the relation
\begin{eqnarray}
\arg (A K_S) \simeq +5.2 \arg (V K_S)
\end{eqnarray}
holds up to a few percent as can be seen in Fig.~\ref{fig:arg}.
For the difference in $\sin 2 \beta$ we obtain
\begin{eqnarray}
|\sin 2\beta_{A K_S}-\sin 2\beta_{V K_S} | \leq 0.19 \cos 2\beta 
 \label{eq:gluinoprediction}
\end{eqnarray}
The correlation with $B \to \phi K_S$ is shown in 
Fig.~\ref{fig:phiccbar} for two values of the 
$\alpha_s$ matrix element of the gluon dipole operator, 
$\kappa_8=-0.045$ (red) corresponding to the asymptotic 
$\phi$ distribution amplitude and $\kappa_8=-0.030$ (green), 
the result in a quark model (see Appendix \ref{app:heff}).
This illustrates the sensitivity to hadronic physics in this NP scenario.
The difference in $\sin 2 \beta$ between axial and vector coupling mesons
is limited if the difference between
$\phi K_S$ and the vector ones is large and negative. 
Current data indicate that
$\sin 2\beta_{A K_S}-\sin 2\beta_{V K_S} \leq 0.04 $.
We checked that this upper bound holds up to a value of the gluon matrix 
element which is 50 percent bigger than our central value $\kappa_8=-0.045$.

With values possible even greater than 60 $ps^{-1}$ for $\Delta m_s$
\cite{Harnik:2002vs}, the
CP asymmetries in the $B_s$-system induced by $\delta^D_{RR \; 23}$ can
dominate over the SM contribution
$15.1 ps^{-1} \leq \Delta m_{s \; SM} \leq 21.0 ps^{-1} $ @ 95 \% C.L. 
\cite{Buras:2002yj}.
Therefore, if at all measurable, this framework allows for large non-SM 
effects in $S_{(c \bar c) \phi}$ and $S_{\phi \phi}$.

\begin{table}[htb]
 \begin{center}
\begin{tabular}{|c|c|c||c|c|}
\hline
$\mbox{}$&$\mbox{both}$ & $Z^\prime_{sb}=0$ &$\mbox{both}$ &$Z^\prime_{sb}=0$\\
\hline
$|\arg(V K_S)| $   &  0.11 & 0.07  & 0.05 & 0.03    \\ 
$|\arg(A K_S)| $   &  0.25 & 0.18  & 0.12 & 0.08    \\ 
$|\sin 2\beta_{A K_S}-\sin 2\beta_{V K_S}| $ & $0.18 \cos 2\beta $& 
$0.12 \cos 2\beta $ & $0.08 \cos 2\beta $ & $0.04 \cos 2\beta $\\
$\sin 2\beta_{\phi K_S}-\sin 2\beta_{V K_S} $ 
& -0.47 ...+0.19 & -0.33...+0.14  & -0.20...+0.14 & -0.12...+0.10 \\
\hline
$|r_Z| $            &  0.50 & 0.16  & 0.19 & 0.07    \\
$|S_{V \phi}| $     &  0.42 & 0.08  & 0.16 & 0.02    \\
$|S_{A \phi}| $     &  0.47 & 0.17  & 0.21 & 0.04    \\
$|S_{\phi \phi}|$   &  0.66 & 0.42  & 0.36 & 0.17    \\
$|S_{A \phi}-S_{V \phi}| $ &  0.15 & 0.10  & 0.07 & 0.05    \\
\hline
\end{tabular}
\caption{ 
Upper bounds on $B_d$ and $B_s$ quantities defined in text 
in the presence of non-standard $Z$-penguins. For
$\sin 2\beta_{\phi K_S}-\sin 2\beta_{V K_S} $ we show the accessible range.
The first two columns are with the current constraints on
the $sZb$ couplings given in Eq.~(\ref{eq:bound}), whereas
in the last two columns we entertain a scenario where
$\sqrt{|Z_{sb}+Z_{sb\; SM}|^2+ |Z_{sb}^\prime|^2} \leq |Z_{sb\; SM}|$, i.e.,
where data on 
${\cal{B}}(b \to s \ell^+ \ell^-)$ show no deviation from the SM.
In the first and third columns both 
helicity couplings are present and the second and fourth ones 
are obtained with $Z^\prime_{sb}=0$.
}\label{tab:Zreach}
 \end{center}
\end{table}

\section{Conclusions  \label{ssix}}

One of todays most precise information on CP violation in the quark sector 
i.e.~$\sin 2 \beta$ given in Eq.~(\ref{beta_psiks})
is an average over several final states with the same 
flavor content. While this procedure returns in the SM 
to very good approximation the same CP-asymmetries, 
it is wrong in general, 
for example if NP breaks parity and CP.

We suggest here to study differences in $\sin 2 \beta$
among decays $B \to (c \bar c) K_S$ into charmonia
$\Psi,\Psi^\prime,\eta_c, \chi_{0,1,2}$
with branching ratios of order $10^{-3}$
to search for physics beyond the SM.
An order one NP effect in $B \to \phi K_S$ decay, which is required
to explain the current anomaly Eq.~(\ref{beta_phiks}) or a similar 
large departure from the SM,  leads to up to ${\cal{O}}(10) \%$ 
in the $B \to (c \bar c) K_S$ amplitude.
This is just at the level at which the data agree with the SM, 
Eqs.~(\ref{beta_psiks}) and (\ref{beta_fit}). 
We stress that differences in $\sin 2 \beta$ can signal NP
independent of improvements in the present error of the CKM fit.
One might extend this program and compare CP asymmetries in decays into
final states with the same flavor other than $c \bar c$.

We have explicitly shown that a NP model with non-standard $Z$-penguins 
does indeed split $\sin 2 \beta$ among
different charmonia, see Eq.~(\ref{eq:Zprediction}).
We stress that  there is a strong correlation 
in this scenario 
between non-SM effects in $b \to c \bar c s$ and $b \to s \bar s s$ decays, 
since both get contributions from the modified 4-Fermi operators.
If the $\phi K_S$ anomaly persists, but 
$\sin 2 \beta_{(J/\Psi, \Psi^\prime) K_S}-\sin 2 \beta_{(\chi_1,\eta_c) K_S}$ 
vanishes, it model independently 
indicates the presence of an enhanced chromomagnetic dipole 
operator \cite{alex} which carries a non-CKM CP violating  phase.
This emphasizes that $b \to c \bar c s$ and $b \to s \bar s s$ decays 
are complementary when constraining and distinguishing NP.
A comparison of $\sin 2 \beta$ differences in the non-SM $Z$-scenario with 
the ones in the MSSM with
gluino mediated FCNCs shows this in Fig.~\ref{fig:phiccbar}.
We discussed time dependent studies in the corresponding decays of 
$B_s$-mesons, i.e.
$B_s \to \phi \phi$ and $B_s \to (c \bar c) \phi$. They show besides 
a similar effect from a NP weak phase on the decay amplitude
one from NP in $B_s \bar B_s$ mixing and
large effects are possible.
The coefficient $S_{c \bar c \phi}$ in the CP asymmetry of 
$B_s \to (c \bar c) \phi$ decays
can be up to ${\cal{O}}(0.4)$ in the presence of $Z$-penguins.
In $B_s \to \phi \phi$ decay they induce an asymmetry
up to order one, i.e.~large as in $B \to \phi K_S$ decay.
Our findings are summarized in Table \ref{tab:Zreach}, where we
also entertain a scenario where the
$b \to s \ell^+ \ell^-$ branching ratio is SM like,
hence the bound on the $Z$-penguins gets tighter 
$\sqrt{|Z_{sb}+Z_{sb\; SM}|^2+ |Z_{sb}^\prime|^2} \leq |Z_{sb\; SM}|$
(assuming zero errors)
than it currently is as given in  Eq.~(\ref{eq:bound}).
There remain sizable effects
in the observables discussed in this work, but closer to potential 
theory SM back grounds.
Information on $Z_{sb}^{(\prime)}$ 
phases also come from the
Forward-Backward asymmetry in $B \to K^* \ell^+ \ell^-$ 
\cite{GGG} and inclusive $B \to X_s \ell^+ \ell^-$ decays in the long term
future.

We studied the effects of non-standard
$sZb$-couplings in an isolated manner. 
If this generic structure arises from a larger model beyond the SM 
e.g.~the MSSM with additional flavor violation beyond CKM,
the signatures can be rather diluted.
We recall that order one mixing between the 
$\tilde c_{L}$ and the $\tilde t_{R}$ squarks 
can generate via chargino-higgsino loops
left handed $sZb$-couplings near the experimental bound, whereas the
right handed couplings are suppressed by the small strange Yukawa \cite{GGG}.
We show for comparison all quantities also 
with flipped helicity $Z_{sb}^\prime$ switched off in Table \ref{tab:Zreach}.
The reach of the $Z$-couplings is sensitive to 
right-handed currents 
in particular in the $B_s$-system, i.e.~$\Delta m_s$.

In our study of CP violating time dependent asymmetries we
neglected NP in transitions between the {\it first} and third generation 
down quarks which could affect $B \bar B$ mixing. Still, 
in this case, differences
in CP asymmetries among different charmonia manifest the presence of
new phases in $b \to c \bar c s$ transitions.
Others places where to look for $Z$-penguins are
$\Lambda_b \to \Lambda \phi$ \cite{Zhao:2002zk},
$\Lambda_b \to \Lambda (\bar c c)$,  $B \to (\bar c c) K^{*}$
and $B \to \phi K^{*}$ decays where polarization observables 
probe the handedness of the NP couplings.
There is experimental support for the possibility of 
large electro weak penguins in $B \to K \pi$ decays \cite{Yoshikawa:2003hb},
which, for example, could be induced by non-standard $Z$-penguins.

\noindent
{\bf Acknowledgments}
G.H.~would like to thank the theory groups at SLAC and LBNL for warm 
hospitality during her stays in the Bay area and Martin Beneke, 
Gerhard Buchalla, Gino Isidori and Alex Kagan for useful discussions.

\begin{appendix}
\renewcommand{\theequation}{\Alph{section}-\arabic{equation}}

\setcounter{equation}{0}
\section{Matrix element of $\bar B \to (c \bar c) K $ and $\bar B \to \phi K$
\label{app:heff}}
The effective Hamiltonian is given as, see e.g.~\cite{BBL}
\begin{eqnarray}
{\cal{H}}_{eff}= -\frac{4 G_F}{\sqrt{2}} V_{tb} V_{ts}^* 
\sum_{i=1\ldots 10, 7 \gamma, 8 g} 
\left( C_i O_i +C^\prime_i O_i^\prime \right)
\end{eqnarray}
with 
\begin{eqnarray}
O_1=( \bar c_{L \alpha } \gamma_\mu b_{L \beta })
( \bar s_{L \beta } \gamma_\mu c_{L \alpha} ) & \mbox{} & ~~~~~~~~~~
O_2=( \bar c_L \gamma_\mu b_L)( \bar s_L \gamma_\mu c_L)  \nonumber \\
O_3=( \bar s_L \gamma_\mu b_L) \sum_{q} \bar q_L \gamma^\mu q_L & \mbox{} 
& ~~~~~~~~~~ 
O_4=( \bar s_{L \alpha } \gamma_\mu b_{L \beta })
\sum_{q} \bar q_{L \beta} \gamma^\mu q_{L \alpha} \nonumber \\
O_5=( \bar s_L \gamma_\mu b_L) \sum_{q} \bar q_R \gamma^\mu q_R & \mbox{} 
& ~~~~~~~~~~ 
O_6=( \bar s_{L \alpha } \gamma_\mu b_{L \beta })
\sum_{q} \bar q_{R \beta} \gamma^\mu q_{R \alpha} \nonumber \\
O_7=\frac{3}{2}( \bar s_L \gamma_\mu b_L) 
\sum_{q} Q_q \bar q_R \gamma^\mu q_R & \mbox{} 
& ~~~~~~~~~~ 
O_8=\frac{3}{2}( \bar s_{L \alpha } \gamma_\mu b_{L \beta })
\sum_{q} Q_q \bar q_{R \beta} \gamma^\mu q_{R \alpha} \nonumber \\
O_9=\frac{3}{2}( \bar s_L \gamma_\mu b_L) 
\sum_{q} Q_q \bar q_L \gamma^\mu q_L & \mbox{} 
& ~~~~~~~~~~ 
O_{10}=\frac{3}{2}( \bar s_{L \alpha } \gamma_\mu b_{L \beta })
\sum_{q} Q_q \bar q_{L \beta} \gamma^\mu q_{L \alpha} \nonumber \\
O_{7 \gamma}= \frac{e}{16 \pi^2} m_b
\bar s_L \sigma_{\mu \nu} b_R F^{\mu \nu}& \mbox{} & ~~~~~~~~~~ 
O_{8 g}=\frac{g_s}{16 \pi^2} m_b
\bar s_{L \alpha} T^a_{\alpha \beta}\sigma_{\mu \nu} b_{R \beta} G^{a \mu \nu}
\label{eq:basis}
\end{eqnarray}
where the sum is over the active quarks $q=u,d,s,c,b$, and $Q_q$ is their
electrical charge in fractions of $e$ and $\alpha, \beta$ are color indices.
The operators $O_i^\prime$ are obtained from flipping 
$L \leftrightarrow R$ in the $O_i$.

The matrix element of decays into mesons with vector coupling such as 
$\bar B \to J/\psi K$ and  $\bar B \to \psi^\prime K$
is proportional to ($N_C$ is the number of colors)
\begin{eqnarray} 
\label{eq:C}
T \!=\! V_{tb} V_{ts}^* \left(\! C_1+C_3 +C_5+\frac{C_2+C_4+C_6}{N_C} +C_7+C_9
+ \frac{C_8+C_{10}}{N_C} \right) \! + C_i \! \to \! C_i^\prime
\end{eqnarray}
The one describing decays into mesons with axial coupling such as
$\bar B \to \chi_1 K$ and $\bar B \to \eta_c K$ can be 
obtained from Eq.~(\ref{eq:C}) by changing 
the sign of 
$C_{5,6,7,8}$
and
$C^\prime_{1,2,3,4,9,10}$.

The matrix element of $\bar B \to \phi K$ is proportional to
\begin{eqnarray} 
P & = & V_{tb} V_{ts}^* \left( C_3+C_4 +C_5+\frac{C_3+C_4+C_6}{N_C} -
\frac{1}{2} (C_7+C_9+C_{10} + \frac{C_8+C_9+C_{10}}{N_C}) \right. \nonumber 
\\ & + & \left.
\kappa_2 C_2 + \kappa_8 
C_{8g}^{eff} \right) + C_i \to C_i^\prime
\label{eq:P}
\end{eqnarray}
where $\kappa_{2,8}$ stem from ${\cal{O}}(\alpha_s)$ corrections 
to the corresponding matrix element 
and the effective coefficient of the chromomagnetic operator, 
$C_{8g}^{eff}$ is defined e.g.~in\cite{BBL}. 
For NP scenarios with an 
enhanced $bs$glue-coupling the matrix element of $O_{8 g}^{(\prime)}$ is 
particularly important. 
Keeping only terms leading in the heavy quark limit, we obtain
\begin{eqnarray}
\label{eq:kappa8}
\kappa_8= - 2 \frac{\alpha_s}{4 \pi} \frac{N_C^2-1}{2 N_C^2} 
\frac{m_b^2}{<q^2>}
= - 2 \frac{\alpha_s}{4 \pi} \frac{N_C^2-1}{2 N_C^2} 
\int_0^1 dx \frac{\Phi_\|(x)}{1-x}
\end{eqnarray}
in agreement with the QCD factorization \cite{Beneke:2001ev} 
calculation by \cite{Cheng:2000hv}.
In the second step we evaluated the averaged momentum square from the 
gluon propagator $1/  \! <q^2 \!>$ 
with  $q^2=m_b^2 (1-x)$ and convoluted it with the longitudinal 
light cone distribution amplitude 
of the $\phi$ meson, $\Phi_\|(x)$, which
encodes the 
momentum distribution of the constituent quarks in the meson,
see \cite{Ball:1998sk} for details.
Here, $x$ denotes 
the fraction of the $\phi$ momentum carried by its $s$-quark.
Theoretical uncertainties in $\kappa_8$ are from $\Phi_\|$
and more generally, from power corrections to the factorization, 
which also limit the accuracy of the total amplitude in Eq.~(\ref{eq:P}).
For asymptotic form $\Phi_\|(x)=6 (1-x) x$, which is supported by 
light cone sum rule calculations \cite{Ball:1998sk} we obtain 
$m_b^2/ \! <q^2 \!>=3$, bigger than the value from a quark model calculation 
$m_b^2/ \! <q^2 \!> \simeq 2$.
For our numerical study we use $\kappa_8=-0.045$ corresponding to the 
asymptotic distribution amplitude 
and $\kappa_2 =-0.011-i 0.012$ \cite{Bander:px}.

\renewcommand{\theequation}{\Alph{section}-\arabic{equation}}

\setcounter{equation}{0}
\section{LLog renormalization for $Z$-penguins \label{app:Zp}}

$Z$-penguins induce the following contributions at the weak scale

\begin{eqnarray}
C_3^Z(m_{weak}) &=& \frac{g^2}{4 \pi^2} \frac{Z_{sb}^*}{V_{tb} V_{ts}^*}
\frac{1}{6} \nonumber \\
C_7^Z(m_{weak}) &=& \frac{g^2}{4 \pi^2} \frac{Z_{sb}^*}{V_{tb} V_{ts}^*}
\frac{2}{3}
\sin^2 \theta_W  \nonumber \\
C_9^Z(m_{weak}) &=& -\frac{g^2}{4 \pi^2} \frac{Z_{sb}^*}{V_{tb} V_{ts}^*}
\frac{2}{3}
(1-\sin^2 \theta_W) \nonumber \\
C_5^{\prime Z}(m_{weak}) &=& \frac{g^2}{4 \pi^2} \frac{Z_{sb}^{\prime
*}}{V_{tb} V_{ts}^*}
\frac{1}{6} \nonumber \\
C_7^{\prime Z}(m_{weak}) &=& -\frac{g^2}{4 \pi^2} \frac{Z_{sb}^{\prime
*}}{V_{tb} V_{ts}^*}
\frac{2}{3}
(1-\sin^2 \theta_W) \nonumber \\
C_9^{\prime Z}(m_{weak}) &=& \frac{g^2}{4 \pi^2} \frac{Z_{sb}^{\prime
*}}{V_{tb} V_{ts}^*}
\frac{2}{3}
\sin^2 \theta_W
\end{eqnarray}

\noindent
Note that non-SM coefficients are treated as
lowest order in $\alpha_s$ and $\alpha_W$ coupling constants.
RG evolution to the $m_b$-scale is done with the effective
12 dimensional LLog anomalous dimension matrix, which is given in  
\cite{BBL}, except for the $\gamma_{ik}$, $i=7,\ldots , 10$ and 
$k=7 \gamma, 8 g$ entries. 
They correspond to the 
$\alpha_s$-mixing of electroweak penguins onto the 
dipole operators in the basis given in Eq.~(\ref{eq:basis}) and 
can be deduced from \cite{misiak} as  ($\gamma=g_s^2/(16 \pi^2) \gamma^0$)
\begin{eqnarray}
\nonumber
\gamma_{7,7 \gamma}^0=-\frac{16}{9} ~~~ \gamma_{7,8 g}^0=\frac{5}{6} ~~~ 
\gamma_{8,7 \gamma}^0=-\frac{1196}{81} ~~~ \gamma_{8,8 g}^0=-\frac{11}{54} \\
\gamma_{9,7 \gamma}^0=\frac{232}{81} ~~~ \gamma_{9,8 g}^0=-\frac{59}{54} ~~~ 
\gamma_{10,7 \gamma}^0=\frac{1180}{81} ~~~ \gamma_{10,8 g}^0=-\frac{46}{27}
\end{eqnarray}
The sectors $O_i$ and $O_i^\prime$ evolve independently with the same
anomalous dimension.
\end{appendix}


\end{document}